\documentclass[proceedings, preprint]{rmaa}

\usepackage{paralist}

\usepackage{psfrag,color}
\usepackage{amsmath}

\SetYear{2013}
\SetConfTitle{Magnetic Fields in the Universe IV}

\title{Magnetohydrodynamic equilibria in barotropic stars} 

\author{
  C. Armaza,\altaffilmark{1} 
  A. Reisenegger,\altaffilmark{1}
  J. A. Valdivia,\altaffilmark{2}
  and P. Marchant\altaffilmark{1,3}}

\altaffiltext{1}{Departamento de Astronom\'ia y Astrof\'isica, Facultad de F\'isica, 
Pontificia Universidad Cat\'olica de Chile, Av. Vicu\~na Mackenna 4860, 782-0436 Macul, 
Santiago, Chile (cyarmaza@uc.cl).}

\altaffiltext{2}{Departamento de F\'isica, Facultad de Ciencias, Universidad de Chile, Casilla 653, Santiago, Chile.}

\altaffiltext{3}{Argelander Institut f\"ur Astronomie, Universit\"at Bonn, Auf dem H\"ugel 71, D-53121, Bonn, Germany.}

\shortauthor{Armaza et al.}
\shorttitle{RevMexAA(SC) Demo Document}

\listofauthors{C. Armaza, A. Reisenegger, J. A. Valdivia, \& P. Marchantr}
\indexauthor{Armaza, C.}
\indexauthor{Reisenegger, A.}
\indexauthor{Valdivia, J. A.}
\indexauthor{Marchant, P.}

\abstract{Although barotropic matter does not constitute a realistic 
model for magnetic stars, it would be interesting to confirm a recent 
conjecture that states that magnetized stars with a barotropic equation of state 
would be dynamically unstable \citep{reisenegger9}. In this work we
construct a set of barotropic equilibria, which can eventually be tested
using a stability criterion. A general description of the ideal MHD 
equations governing these equilibria is summarized, allowing for both
 poloidal and toroidal magnetic field components. A new finite-difference 
numerical code is developed in order to solve the so-called Grad-Shafranov 
equation describing the equilibrium of these configurations, and some 
properties of the equilibria obtained are briefly discussed.}

\resumen{Aunque la materia barotr\'opica no constituye un modelo realista
para estrellas magn\'eticas, ser\'ia interesante confirmar una conjetura 
reciente que establece que las estrellas magn\'eticas con ecuaci\'on de estado
barotr\'opica, ser\'ian din\'amicamente inestables \citep{reisenegger9}.
En este trabajo construimos un conjunto de equilibrios barotr\'opicos, los cuales
pueden ser finalmente testeados usando un criterio de estabilidad. Una descripci\'on
general de las ecuaciones de MHD ideal que gobiernan estos equilibrios es
revisada, permitiendo tanto una componente poloidal, como una componente
toroidal del campo magn\'etico. Un nuevo c\'odigo num\'erico en diferencia 
finita es desarrollado para resolver la llamada ecuaci\'on de Grad-Shafranov
que describe el equilibrio de estas configuraciones, y algunas propiedades de los
equilibrios obtenidos son brevemente discutidas.}

\addkeyword{Stars: MHD}
\addkeyword{Stars: barotropic}

\begin{document}
\maketitle

\section{Baro... what?}
\label{intro}

Barotropic equations of state, where pressure is a function solely of density,
are often assumed to describe the matter within magnetic stars in ideal
magnetohydrodynamic (MHD) equilibrium \citep{yoshida6, haskell8, lander9}. 
Barotropy strongly restricts the range of possible equilibrium configurations,
and does not strictly represent the realistic stably stratified matter within
these objects, which is likely to be an essential ingredient in
the stability of magnetic fields in stars \citep{reisenegger9}. With this
 in mind, it is interesting to carry out the pedagogical exercise of checking 
whether the unrealistic barotropic equilibria are really stable or not. This work
is focused on obtaining a wide range of these equilibria and study their main
properties, as a starting point to study their stability.\\

\section{MHD equilibria: the Grad-Shafranov equation}

\begin{figure}
\begin{center}
\includegraphics[width=4cm]{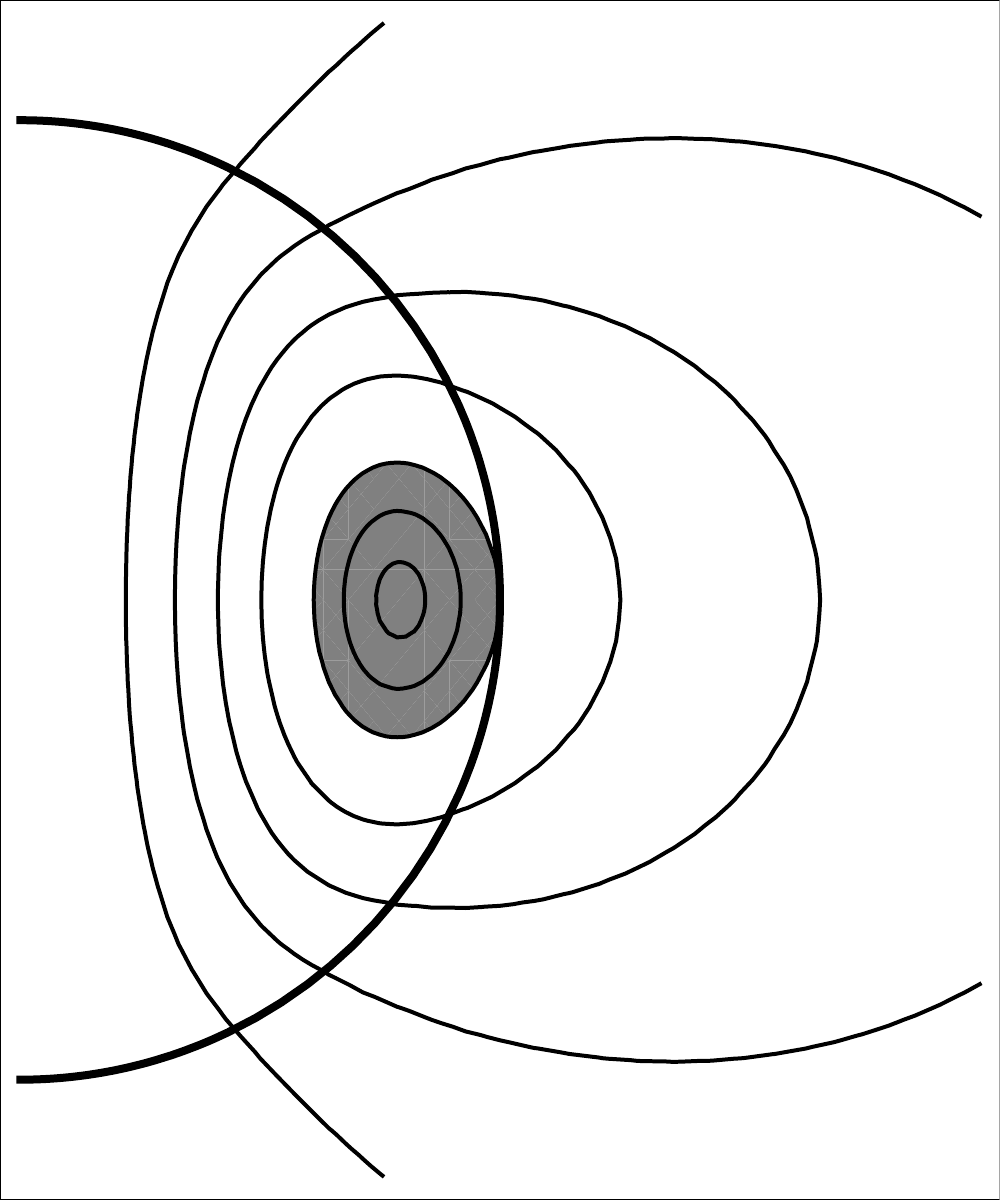}
\caption{Meridional cut of a star bearing an axisymmetric magnetic field. 
The bold curve is the surface of the star, while the thinner curves are poloidal 
field lines. The toroidal component of the magnetic field may lie only in regions
 where the poloidal field lines close inside the star (gray region).}\label{mustlie}
\end{center}
\end{figure}

In the ideal MHD approximation, a magnetic star may be considered as a perfectly conducting
fluid in dynamical equilibrium described by the Euler equation,
\begin{equation}
\boldsymbol\nabla P + \rho\boldsymbol\nabla\Phi = \frac{1}{c}\mathbf J \times \mathbf B,
\end{equation} 
where the right-side is the Lorentz force per unit volume. Considered objects have a very large
fluid pressure $P$ ($P\sim GM^2/R^4$, $M$ being the mass and $R$ the radius),
 to magnetic pressure $B^2/8\pi$ ratio ($B$ being a characteristic magnetic field strength), 
$8\pi P/B^2 \sim 10^6$  \citep{reisenegger9}, which suggests that magnetic forces 
may be balanced by a slight perturbation of an unmagnetized spherical background equilibrium.
This implies that, as an approximation, we can consider the star as spherical with negligible 
deformations due to the magnetic forces. In addition, if axial symmetry is assumed, and 
spherical coordinates $(r,\theta,\phi)$ are used to describe the model, all scalar quantities 
are independent of the azimuthal coordinate, and the magnetic field may be expressed
 as the sum of a \textit{poloidal} (meridional field lines) component, and a 
\textit{toroidal} (azimuthal field lines) component, each determined by a single 
scalar function,
\begin{equation}
\mathbf B = \mathbf B_\text{pol} + \mathbf B_\text{tor} = \boldsymbol\nabla\alpha(r,\theta) \times
 \boldsymbol\nabla\phi + \beta(r,\theta)\boldsymbol\nabla\phi,
\end{equation}
which turn out to be constant along their respective field lines \citep{chandrasekhar56}. 
Under this symmetry, the azimuthal component of the magnetic force per unit volume must 
vanish, which implies a functional relation between these scalar functions, 
$\beta(r,\theta) = \beta\left(\alpha(r,\theta)\right)$. In this way, both $\alpha$ and
 $\beta$ are constant along field lines and, if vacuum is assumed outside the star, 
the toroidal field may lie only in regions where the poloidal field lines close within 
the star (Figure \ref{mustlie}).
On the other hand, if a barotropic equation of state, $P=P(\rho)$, is assumed, 
the Lorentz force per unit mass must be the gradient of some arbitrary function 
$\chi(r,\theta)$, which turns out to be a function of $\alpha$ as well, 
$\chi(r,\theta) = \chi\left(\alpha(r,\theta)\right)$. Using all this formalism,
 a non-linear elliptic partial differential equation is found to be the master
 equation governing the equilibrium of a barotropic MHD equilibrium, 
the so-called Grad-Shafranov (GS) equation,
\begin{equation}
\frac{\partial^2\alpha}{\partial r^2} + 
\frac{\sin\theta}{r^2}\frac{\partial}{\partial\theta}\left(\frac{1}{\sin\theta}\frac{\partial\alpha}{\partial\theta}\right)
+ \beta\beta' + r^2\sin^2\theta\rho \chi' = 0
\end{equation}
\citep{grad58, shafranov66} where primes stand for derivative with respect 
to the argument, and both $\beta=\beta(\alpha)$ and $\chi=\chi(\alpha)$ are 
two arbitrary functions, whose form may be chosen depending on the particular 
magnetic configuration of interest. Under the assumption of weak magnetic field
discussed in \S\ref{intro}, the density $\rho$ appearing in the GS equation may
be replaced by its non-magnetic background counterpart, $\rho=\rho(r)$, such 
that we solve for the magnetic functions for a \emph{given} density profile, instead
of considering the more difficult problem of solving self-consistently 
for the magnetic functions \emph{and} for the fluid quantities.  

\section{Numerical solutions}
 
\begin{figure}[ht]
\includegraphics[width=3.93cm]{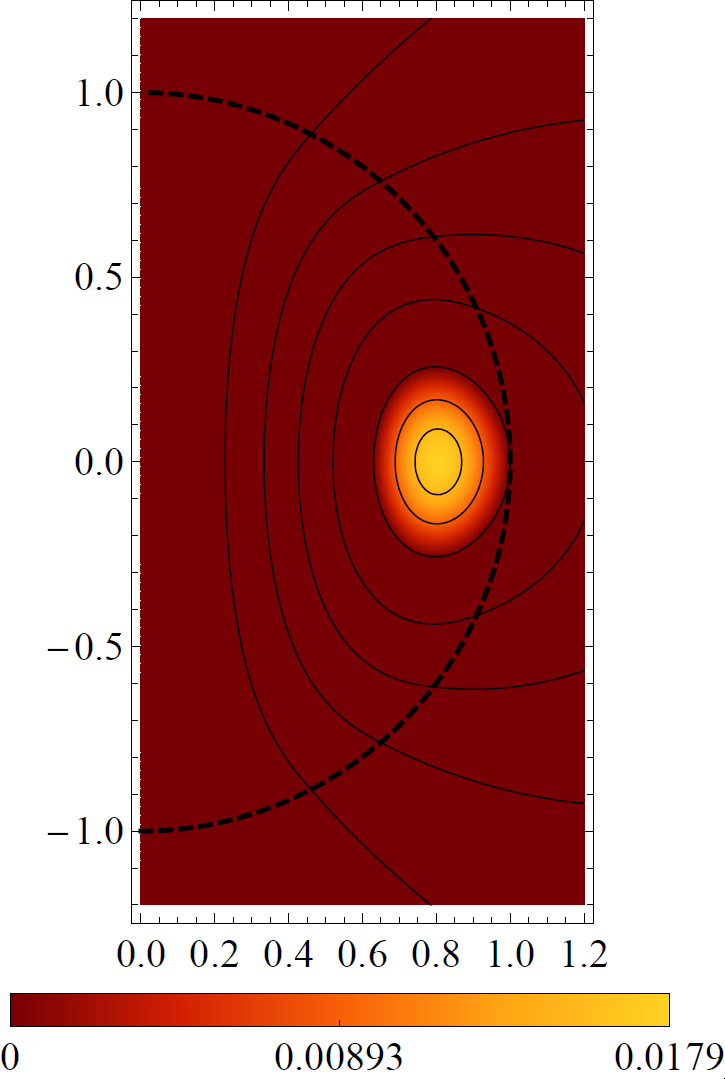}$\;$\includegraphics[width=3.93cm]{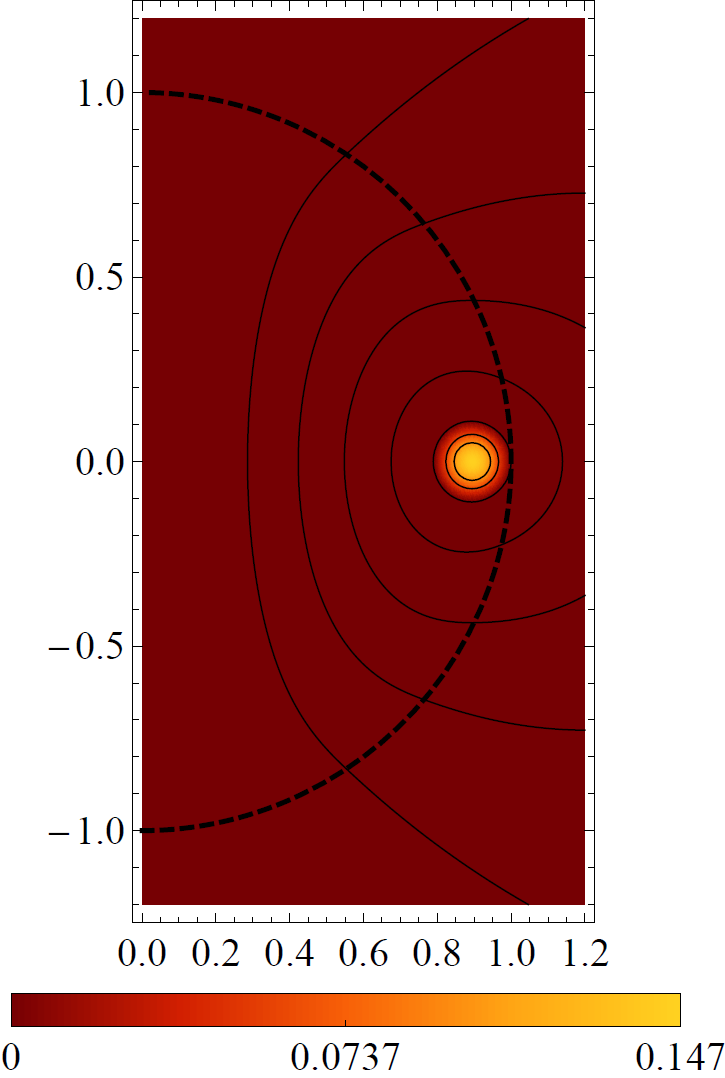}
\caption{Numerical equilibria found with our code. Left: $s=10$, with $E_\text{tor}/E_\text{mag}
 \approx 0.5\%{}$. Right: $s=35$, with $E_\text{tor}/E_\text{mag}
 \approx 3.2\%{}$.}\label{results}
\end{figure}

\begin{figure}[ht]
\includegraphics[width=7.8cm]{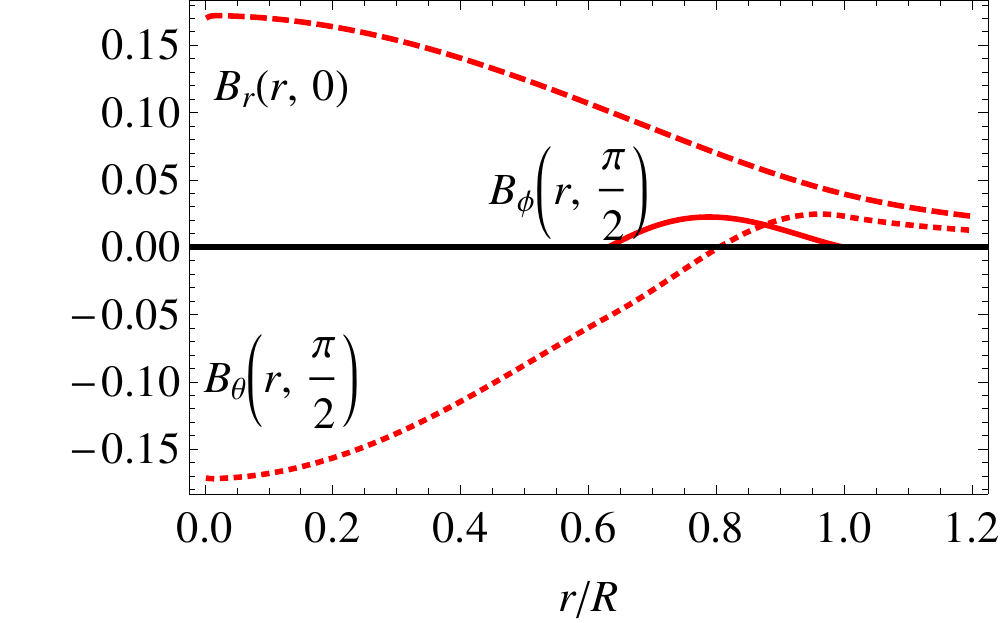}
\caption{Magnitude of the magnetic field for the equilibria shown in Figure 
\ref{results} with $s=10$. The maximum toroidal strength is about one order of magnitude
smaller than the poloidal one.}\label{s10}
\end{figure}

\begin{figure}[ht]
\includegraphics[width=7.8cm]{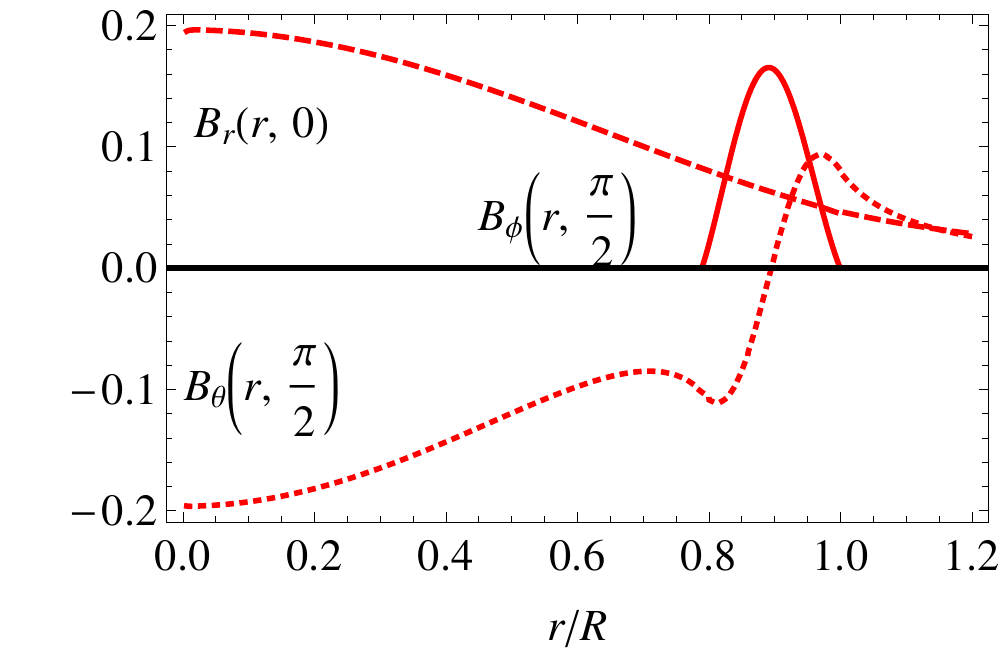}
\caption{Magnitude of the magnetic field for the equilibria shown in Figure 
\ref{results} with $s=35$. Both the poloidal and the toroidal maximum strength
are of the same order of magnitude.}\label{s35}
\end{figure}

Outside the star, $\alpha$ corresponds to an infinite superposition of 
multipoles, which is the general solution of the GS equation with both $\beta = 0$ 
and $\rho=0$. We have implemented a finite-difference code to solve numerically 
the GS equation inside the star, for arbitrary choices of $\beta(\alpha)$ 
and $\chi(\alpha)$. Solutions are matched to the exterior expansion by 
demanding continuity of $\alpha$ and its derivatives (related to the 
magnetic field components), in order to avoid surface currents. After
testing our code, we obtained barotropic equilibria for the particular case
$\chi(\alpha) = \alpha$, $\rho(r) = \rho_c(1-r^2/R^2)$ and
\begin{equation}\label{beta}
\beta(\alpha) = \begin{cases}
s(\alpha-\alpha_s)^{1.1} & \alpha_s \leq \alpha\\
0 & \alpha < \alpha_s,
\end{cases}
\end{equation}
where $s$ is a free parameter accounting for the relative strength between 
the poloidal and the toroidal component. In the definition below, an 
exponent larger than 1 was chosen in order to prevent a discontinuous 
$\beta'$ at the layer between regions with and without toroidal field; it 
is found that larger exponents than 1.1 give a smaller toroidal field strength. Also,
 $\alpha_s\equiv \alpha(R,\pi/2)$ stands for the value of $\alpha$ along the
 largest poloidal field line closing within the star, being $R$ the 
stellar radius, so the toroidal field lies in the region inside the
curve $\alpha_s = \alpha(r,\theta)$ only. Figure \ref{results} shows some numerical results for 
the particular $\beta(\alpha)$ in Eq. (\ref{beta}). Black lines correspond 
to poloidal field lines with $0.2\alpha_s$, $0.4\alpha_s$, $0.6\alpha_s$,
$0.8\alpha_s$, $1.0\alpha_s$, $1.08\alpha_s$ and $1.13\alpha_s$, respectively,
whereas the color map accounts for the strength of the toroidal field. 
In turn, Figures \ref{s10}-\ref{s35} show the strength of the magnetic 
field along the axis and the equatorial line for these equilibria.\\


\section{Discussion}

All the equilibria found consist of a mixed poloidal-toroidal field with a dominant 
poloidal component in the magnetic energy $E_\text{mag}$. For the cases 
studied so far, the energy stored in the toroidal component $E_\text{tor}$ 
is only a few percent of the total magnetic energy, even in cases where the 
maximum strength of the toroidal field is comparable to that of the poloidal component:
 the larger the toroidal field, the smaller the volume where it lies. 
This small contribution to the energy has already been reported in the literature,
 but assuming a purely dipolar magnetic field outside the star \citep{lander12}. 
Our code, allowing an arbitrary number of multipoles, seems to indicate that 
higher multipoles do not contribute significantly to the energy of these equilibria, 
at least not for small to moderate values of $s$. It is desirable to study this fact 
in more detail and confirm, for instance, whether a global maximum for 
$E_\text{tor}/E_\text{mag}$ exists, already reported using general-relativistic MHD \citep{ciolfi9}.
 Once we obtain a wide range of relevant equilibria with consistent physical choices of the arbitrary 
magnetic functions, their dynamical stability may be analyzed using either 
a perturbative analysis or numerically solving the time-evolution of such configurations.\\

\section{Acknowledgments}

All authors are supported by CONICYT International Collaboration Grant DFG-06.
CA and PM are supported by a CONICYT Master's Fellowship. AR and JAV are supported by
FONDECYT Regular Grants 1110213 and 1110135, respectively. CA, AR and PM are 
supported by the Basal Center for Astrophysics and Associated Technologies.

\end{document}